\documentclass[english,aps,prc,nofootinbib,superscriptaddress,twocolumn,letter]{revtex4}
\usepackage[latin9]{inputenc}
\usepackage{hyperref}
\hypersetup{
  colorlinks=true,        % false: boxed links; true: colored links
  linkcolor=blue,         % color of internal links
  citecolor=cyan,         % color of links to bibliography
}
\usepackage{breakurl}
\usepackage{graphicx}
\usepackage{epsf}
\usepackage{epsfig}
\usepackage{amssymb,amsmath}
\usepackage[usenames]{color}
\usepackage{amssymb}
\usepackage{times}
\usepackage{comment}
\usepackage{xfrac}

\allowdisplaybreaks



\begin{document}

\preprint{This line only printed with preprint option}

\title{System size and shape dependences of collective flow fluctuations in relativistic nuclear collisions}

\author{Xinrong Chen}
\affiliation{Institute of Frontier and Interdisciplinary Science, Shandong University, Qingdao, Shandong, 266237, China}

\author{Xiang-Yu Wu}
\email{xiangyu.wu2@mail.mcgill.ca}
\affiliation{Institute of Particle Physics and Key Laboratory of Quark and Lepton Physics (MOE), Central China Normal University, Wuhan, Hubei, 430079, China}
\affiliation{Department of Physics, McGill University, 3600 University Street, Montreal, QC, Canada H3A 2T8}
\affiliation{Department of Physics and Astronomy, Wayne State University, Detroit MI 48201.}

\author{Shanshan Cao}
\email{shanshan.cao@sdu.edu.cn}
\affiliation{Institute of Frontier and Interdisciplinary Science, Shandong University, Qingdao, Shandong, 266237, China}

\author{Guang-You Qin}
\email{guangyou.qin@ccnu.edu.cn}
\affiliation{Institute of Particle Physics and Key Laboratory of Quark and Lepton Physics (MOE), Central China Normal University, Wuhan, Hubei, 430079, China}

\begin{abstract}
Quantum fluctuations plays an essential role in forming the collective flow of hadrons observed in relativistic heavy-ion collisions. Event-by-event fluctuations of the collective flow can arise from various sources, such as the fluctuations in the initial geometry, hydrodynamic expansion, hadronization, and hadronic evolution of the nuclear matter, while the exact contribution from each source is still an open question. Using a (3+1)-dimensional relativistic hydrodynamic model coupled to a Monte-Carlo Glauber initial condition, Cooper-Frye particlization and a hadronic transport model, we explore the system size and shape dependences of the collective flow fluctuations in Au+Au, Cu+Au, and O+O collisions at $\sqrt{s_\mathrm{NN}}=200$~GeV. The particle yields, mean transverse momenta, 2-particle and 4-particle cumulant elliptic flows ($v_2\{2\}$ and $v_2\{4\}$) from our calculation agree with the currently existing data from RHIC. Different centrality dependences of the flow fluctuations, quantified by the $v_2\{4\}/v_2\{2\}$ ratio, are found for different collision systems due to their different sizes and shapes. By comparing $v_2\{4\}/v_2\{2\}$ between different hadron species, and comparing $v_2\{4\}/v_2\{2\}$ to the initial state geometric fluctuations quantified by the cumulant eccentricity ratio $\varepsilon_2\{4\}/\varepsilon_2\{2\}$, we find that while the initial state fluctuations are the main source of the $v_2$ fluctuations in large collision systems, other sources like nonlinear hydrodynamic response, hadronization, and hadronic afterburner can significantly affect the $v_2$ fluctuations in small systems.
\end{abstract}

\maketitle
\date{\today}

\section{Introduction}
\label{section1}

Extensive research conducted at the Relativistic Heavy-Ion Collider (RHIC) and the Large Hadron Collider (LHC) suggests that a color deconfined state of QCD matter, known as quark-gluon plasma (QGP), is produced at extraordinarily high temperature and density. Ample evidence from experiments~\cite{STAR:2003xyj,ALICE:2010suc,ATLAS:2014txd,CMS:2012zex} indicates the QGP is a strongly coupled matter that behaves like a nearly perfect fluid and possesses the smallest specific shear viscosity one has ever achieved in laboratory~\cite{Adams:2012th,Bernhard:2019bmu}. The most notable fluid property of the QGP is its collectivity, which can be quantified by the collective flow coefficient $v_n$, or the $n$-th order Fourier coefficient of the azimuthal angular distribution of particles emitted from the QGP~\cite{Voloshin:1994mz,Poskanzer:1998yz}:
\begin{equation}
    \frac{dN}{d\phi} = \frac{N}{2\pi} \left\{ 1 + 2  \sum_{n=1}^\infty  v_n \cos\left[n (\phi - \Psi_n) \right] \right\},
\label{eq:1}
\end{equation}
where $\phi$ represents the azimuthal angle in the plane transverse to the beam direction, and $\Psi_n$ is the $n$-th order event plane angle which maximizes the value of $v_n$.
Over the past two decades, considerable efforts have been devoted to developing relativistic hydrodynamics models, which have now become a highly successful and therefore the standard model for describing particle production and their collective flow coefficients observed in high-energy nuclear collisions~\cite{Heinz:2013th,Gale:2013da,Luzum:2008cw,Romatschke:2017ejr,Rischke:1995ir,Huovinen:2013wma,Pang:2012he,Wu:2018cpc,Zhao:2017yhj,Pang:2018zzo,Ding:2021ajz,Ze-Fang:2017ppe,Shen:2014vra,Jiang:2021ajc}.

While the second order (elliptic) flow $v_2$ is mainly driven by the average elliptic geometry of the overlapping region between the projectile and target nuclei in non-central collisions, higher order flow coefficients can arise from initial state fluctuations that generates triangular, quadrangular, pentagonal and even higher order geometric components of the overlapping region. Apart from the initial state, fluctuations also exists in hydrodynamic expansion, hadronization, and hadronic rescatterings. Charting contribution from each of these sources to the flow fluctuations in the final state is still an ongoing effort. To quantify the event-by-event fluctuations of collective flow coefficients, one may evaluate these coefficients using the azimuthal correlations between final state particles, such as 2-, 4-, and 6-particle correlations, instead of Eq.~(\ref{eq:1}). The cumulant method~\cite{Poskanzer:1998yz,Borghini:2000sa,Borghini:2001zr} suggests the flow coefficients evaluated from different orders of cumulants, or correlations between different numbers of particles, yield different values; and the discrepancies between these values directly reflect the fluctuations of collective flows. Another advantage of this cumulant method is to avoid the difficulty in determining the event plane angle $\Psi_n$ in Eq.~(\ref{eq:1}) in realistic measurements.

Flow fluctuations in large nuclear collision systems (Pb+Pb and Au+Au) at the LHC and RHIC energies has been widely explored by both experimental measurements~\cite{Magdy:2018itt,STAR:2022gki} and theoretical calculations~\cite{Alba:2017hhe,Schenke:2019ruo,Magdy:2020gxf,Magdy:2021sba,Rao:2019vgy,Wu:2021fjf,Zhu:2024tns}. In these sufficiently large systems, it has been found that the flow fluctuations is not sensitive to the collision energy ($\sqrt{s_\mathrm{NN}}$); instead, it is considerably influenced by the geometrical fluctuations in the initial state, especially in central collisions.
Similar studies have also been extended to other collision systems with various sizes~\cite{Sievert:2019zjr,Schenke:2020mbo}.
An additional way to investigate the correlation between the initial state geometry and the final state collective flow is through asymmetrical collisions between different nuclei, such as the Cu+Au collisions measured by both 
PHENIX~\cite{PHENIX:2015zbc} and STAR~\cite{STAR:2017ykf,STAR:2022gki} Collaborations at RHIC. Related theoretical explorations have also been conducted based on the AMPT simulation~\cite{Chen:2005zy,He:2020xps}. 

For a timely understanding of the recent flow data from Cu+Au collisions~\cite{STAR:2022gki} and the undergoing measurement on O+O collisions at RHIC~\cite{Huang:2023viw}, in this work, we systematically compare the collective flow fluctuations between Au+Au, Cu+Au, and O+O collisions at $\sqrt{s_\mathrm{NN}}=200$~GeV using the (3+1)-dimensional (D) viscous hydrodynamic model CLVisc~\cite{Pang:2018zzo,Wu:2021fjf}. By fitting the hydrodynamic model parameters to existing data in Au+Au collisions, we calculate the identified particle yields, their mean transverse momenta ($p_\mathrm{T}$), and collective flow coefficients from multi-particle correlations across the three systems above.  
From the ratio of $v_2$ between 4 (6)-particle correlation and 2-particle correlation methods ($v_2\{4\}/v_2\{2\}$ and $v_2\{6\}/v_2\{2\}$), we observe the flow fluctuations is sensitive to both the system size and the medium geometry in the initial state. To further explore the source of these fluctuations, the ratios of the cumulant eccentricities~\cite{Qiu:2011iv,Ma:2016hkg} of the initial medium geometry ($\varepsilon_2\{4\}/\varepsilon_2\{2\}$ and $\varepsilon_2\{6\}/\varepsilon_2\{2\}$) are evaluated and compared to the cumulant $v_2$ ratios, which provides a direct way to illustrate the conversion of the initial geometric fluctuations into the final collective flow fluctuations. 

The rest of this paper is organized as follows. Section~\ref{section2} describes the theoretical framework used in this work, including the Monte-Carlo Glauber model for the initial condition, the CLVisc hydrodynamic simulation for the QGP expansion, the Cooper-Frye formalism for hadronization, and the hadronic transport model SMASH for rescatterings between hadrons. Section~\ref{section3} compares our numerical results on the particle yields, mean $p_\mathrm{T}$, and cumulant collective flow coefficients of both charged and identified hadrons between Au+Au, Cu+Au, and O+O collisions at $\sqrt{s_\mathrm{NN}}=200$~GeV. In the end, a summary and outlook is presented in Section~\ref{section4}.

%%%%% Sec 2
\section{Framework}
\label{section2}

In this section, we present the MC-Glauber model for initializing the entropy and net baryon density distributions of the QGP, the (3+1)-D viscous hydrodynamic model CLVisc for event-by-event simulation of the QGP expansion, the Cooper-Frye formalism for converting the QGP into hadrons and the SMASH model for hadronic rescatterings, together with the model parameters we use in this work.

\subsection{Initial condition}
\label{section2-1}

We start with a three-parameter parametrization of the nucleon density distribution function inside a nucleus as~\cite{DeVries:1987atn},
\begin{equation}
    \rho(r_p) =  \left(1 + w\frac{r_p^2}{R^2}\right) \frac{\rho_0}{1 + \exp\left(\frac{r_p-R}{a}\right)},
\label{eq:3}
\end{equation}
where $r_p$ represents the radial position of a nucleon, $\rho_0=0.17$~fm$^{-3}$ is the equilibrium density of nuclear matter, $R$ and $a$ are the radius and the surface thickness parameter of the nucleus. The equation above returns to the standard two-parameter Woods-Saxon distribution with $w=0$. The model parameters of $^{197}$Au, $^{64}$Cu, and $^{16}$O nuclei used in our present study are listed in Tab.~\ref{tab:table1}. 
According to these distributions, we use the Monte-Carlo (MC) method to sample the positions of nucleons inside the projectile and target nuclei. To prevent two nucleons from being too close to each other inside a nucleus, a minimum distance of $d = 0.81$~fm is imposed in our MC sampling.

\begin{table}[t]
\caption{\label{tab:table1}%
Parameters for nucleon density distributions in different nuclei.
}
\begin{ruledtabular}
\begin{tabular}{cccc}
\textrm{}&
\textrm{$R$ (fm)}&
\textrm{$a$ (fm)}&
\textrm{$w$}\\
\colrule
$^{197}$Au & 6.38 & 0.535 & 0\\
$^{64}$Cu & 4.21 & 0.598 & 0\\
$^{16}$O & 2.61 & 0.513 & $-0.051$\\
\end{tabular}
\end{ruledtabular}
\end{table}

By assuming high-energy nucleons stream along straight lines without changing their directions when being scattered, we let two nucleons (one from projectile along the $+\hat{z}$ direction and one from target along the $-\hat{z}$ direction) collide when their transverse distance is smaller than $(\sigma_\mathrm{NN}^\mathrm{inel}/\pi)^{-\frac{1}{2}}$, where the inelastic nucleon-nucleon cross section is taken as $\sigma_\mathrm{NN}^\mathrm{inel}=42$~mb at $\sqrt{s_\mathrm{NN}}=200$~GeV~\cite{PHOBOS:2007vdf}. The positions of both binary collisions (taken as the mid-points of nucleon pairs) and nucleons that participate in these collisions (or participant nucleons) are recorded and contribute to particle production from a nucleus-nucleus collision event. The multiplicity of the final state charged particles then follow the form~\cite{Qin:2010pf}
\begin{equation}
    N_\mathrm{ch} \propto \alpha N_\mathrm{part} + ( 1- \alpha)  N_\mathrm{bin},
\label{eq:4}
\end{equation}
where $\alpha$ is a parameter controlling the relative contribution from the participant nucleon number ($N_\mathrm{part}$) and the binary collision number ($N_\mathrm{bin}$), which can be determined by the centrality dependence of charged particle yield later. We note that although the contribution from binary collisions might be negligible when the nuclear collisions are less energetic, e.g. $\sqrt{s_\mathrm{NN}} \lesssim 62.4$~GeV~\cite{Wu:2021fjf}, it is crucial for a simultaneous description of hadron observables at different centralities at $\sqrt{s_\mathrm{NN}}=200$~GeV in our present work. Although this two-component Glauber model has been shown inaccurate in ultra-central collisions~\cite{STAR:2015mki}, it is still considered a reasonable and convenient model for the initial condition of QGP as long as the centrality is not very small.

\begin{table*}[tbp!]
\caption{\label{tab:table2}%
Parameters for constructing the 3-dimensional initial entropy density and normalized baryon number density distributions for hydrodynamic evolution at $\sqrt{s_\mathrm{NN}}= 200$~GeV.
}
\begin{ruledtabular}
\begin{tabular}{cccccccccccc}
\textrm{$\alpha$} &
\textrm{$K$} &
\textrm{$\tau_0$ (fm/$c$)} &
\textrm{$\sigma_r$ (fm)} &
\textrm{$\sigma_s$} &
\textrm{$\eta_0^\mathrm{s}$} &
\textrm{$\sigma_{n;\mathrm{P}}$} &
\textrm{$\sigma_{n;\mathrm{T}}$} &
\textrm{$\eta_{0;\mathrm{P}}^n$} &
\textrm{$\eta_{0;\mathrm{T}}^n$} &
\textrm{$\sigma_\mathrm{w}$} &
\textrm{$\eta_\mathrm{w}$} \\
\colrule
0.8 & 11.55 & 0.75 & 0.5 & 0.65 & 2.5 & 0.07 & 1.4 & 3.5 & $-3.5$ &  1.2 & 1.0 \\
\end{tabular}
\end{ruledtabular}
\end{table*}

The local entropy density $s$ and the normalized baryon number density $n_0$ then read:
\begin{align}
&s(x,y,\eta_\mathrm{s}) = \alpha s_\mathrm{part}(x,y,\eta_\mathrm{s}) + (1-\alpha) s_\mathrm{bin}(x,y,\eta_\mathrm{s}), \label{eq:s3D} \\
&n_0(x,y,\eta_\mathrm{s}) =  n_\mathrm{part} (x,y,\eta_\mathrm{s}),
\label{eq:5-6}   
\end{align}
where the former is contributed by both participant nucleons and binary collisions, while the latter is only contributed by participant nucleons. Here, $x$ and $y$ are coordinates in the transverse plane, and $\eta_\mathrm{s}$ denotes the spacetime rapidity.
Following Ref.~\cite{Wu:2021fjf}, $s_\mathrm{part}$ and $n_\mathrm{part}$ above at the initial time ($\tau_0$) of hydrodynamic evolution are given by
\begin{align}
    &s_\mathrm{part} = \frac{K}{\tau_0} \left[H_\mathrm{P}^s(\eta_\mathrm{s})\tilde{s}_\mathrm{P}(x,y) + H_\mathrm{T}^s(\eta_\mathrm{s})\tilde{s}_\mathrm{T}(x,y)\right],\\
    &n_\mathrm{part} = \frac{1}{\tau_0} \left[H_\mathrm{P}^n(\eta_\mathrm{s})\tilde{s}_\mathrm{P}(x,y) + H_\mathrm{T}^n(\eta_\mathrm{s})\tilde{s}_\mathrm{T}(x,y)\right],
\label{eq:7-8}
\end{align}
where $\tilde{s}_\mathrm{P}$ and $\tilde{s}_\mathrm{T}$ represent entropy density in the transverse plane contributed by the projectile and target respectively, $H_\mathrm{P/T}^s$ and $H_\mathrm{P/T}^n$ are the longitudinal envelop functions for entropy density and baryon number density respectively, and $K$ is an overall factor that controls the magnitude of the initial entropy density. 

The $\tilde{s}_\mathrm{P/T}$ function can be constructed using the distribution of participant nucleons in the projectile/target nucleus as 
\begin{equation}
    \tilde{s}_\mathrm{P/T} = \sum_{i\in\mathrm{P/T}} \frac{1}{2\pi\sigma_r^2}\exp\left[\frac{-(x - x_i)^2 - (y - y_i)^2}{2\pi\sigma_r^2}\right],
\end{equation}
where $(x_i,y_i)$ represents the transverse position of the $i$-th participant nucleon obtained from the MC-Glauber model, $\sigma_r$ is the transverse Gaussian smearing width. The longitudinal envelop functions take the forms of~\cite{Bozek:2010bi,Bozek:2011if,Bozek:2013uha,Denicol:2018wdp,Wu:2021fjf}  
\begin{align}
H^s_\mathrm{P/T} &= \theta(\eta_\mathrm{max} - \lvert\eta_\mathrm{s}\rvert) \left(1 \pm \frac{\eta_\mathrm{s}}{y_\mathrm{beam}}\right) \Biggl[ \theta(\eta_0^\mathrm{s}- \lvert\eta_\mathrm{s}\rvert)  \nonumber\\
& + \theta( \lvert\eta_\mathrm{s}\rvert - \eta_0^\mathrm{s}) \exp \left( \frac{(\lvert\eta_\mathrm{s}\rvert - \eta_0^\mathrm{s})^2}{2\sigma_s^2}\right) \Biggr], \\
H^n_\mathrm{P/T} &= \frac{1}{N} \Biggl[ \theta(\eta_\mathrm{s} - \eta^n_{0;\mathrm{P/T}}) \exp \left( - \frac{(\eta_\mathrm{s} - \eta^n_{0;\mathrm{P/T}})^2}{2\sigma_{n;\mathrm{P/T}}^2}\right)  \nonumber\\
& + \theta(\eta^n_{0;\mathrm{P/T}} - \eta_\mathrm{s}) \exp \left( - \frac{(\eta_\mathrm{s} - \eta^n_{0;\mathrm{P/T}})^2}{2\sigma_{n;\mathrm{T/P}}^2}\right) \Biggr],
\end{align}
in which parameters $\eta_0^\mathrm{s}$, $\sigma_s$, $\eta^0_{n;\mathrm{P/T}}$, $\sigma_{n;\mathrm{T/P}}$ will be determined by the charged particle distribution along the longitudinal direction, $N$ is the normalization factor for $n_0$, and $\eta_\mathrm{max}=y_\mathrm{beam}=\mathrm{arctanh}(v_\mathrm{beam})$ is the rapidity of a projectile beam nucleon, or the maximum $\eta_\mathrm{s}$ it can reach with velocity $v_\mathrm{beam}=\sqrt{s_\mathrm{NN}}/(2m_\mathrm{N})$ with $m_\mathrm{N}$ the nucleon mass. Alternative ways of introducing the longitudinal profiles of the QGP are discussed in Refs.~\cite{Hirano:2005xf,Okai:2017ofp}.

Similarly, one can define the $s_\mathrm{bin}$ part in Eq.~(\ref{eq:s3D}) as 
\begin{equation}
    s_\mathrm{bin} = \frac{K}{\tau_0} H^s_\mathrm{bin}(\eta_\mathrm{s})\tilde{s}_\mathrm{bin}(x,y),
\label{eq:9}
\end{equation}
with 
\begin{align}
    &\tilde{s}_\mathrm{bin} = \sum_{i\in\mathrm{bin}} \frac{1}{2\pi\sigma_r^2}\exp\left[\frac{-(x - x_i)^2 - (y - y_i)^2}{2\pi\sigma_r^2}\right],\label{eq:sbin}\\
    &H^s_\mathrm{bin} = \exp\left[ - \frac{(\eta_\mathrm{s}-\eta_\mathrm{w})^2}{2\sigma_\mathrm{w}^2}  \theta(\lvert\eta_\mathrm{s}\rvert - \eta_\mathrm{w}) \right],
\label{eq:Hsbin}
\end{align}
where $i$ runs over the binary collision points in a nucleus-nucleus collision event, $\eta_\mathrm{w}$ and $\sigma_\mathrm{w}$ are model parameters.

We summarize in Tab.~\ref{tab:table2} the parameters we use for constructing the initial entropy density and the baryon number density through Eq.~(\ref{eq:s3D}) to Eq.~(\ref{eq:Hsbin}). They are fitted from the existing data in Au+Au collisions at $\sqrt{s_\mathrm{NN}}=200$~GeV in the next section. We assume these parameters are solely sensitive to the beam energy, and remain the same across different collision systems. The evolution of nuclear matter prior to the QGP phase is not included in this work, which becomes more important for lower energy collisions~\cite{Dore:2020fiq}. To partly compensate effects of this pre-equilibrium evolution, we choose the initial proper time to be larger than the overlap time between the colliding nuclei, $\tau_0>2R/\sinh(y_\mathrm{beam})$, in order to allow more time for the system to approach local equilibrium before hydrodynamics starts.

\begin{figure*}[tbp!]
  \centering
  \includegraphics[width=0.96\textwidth]{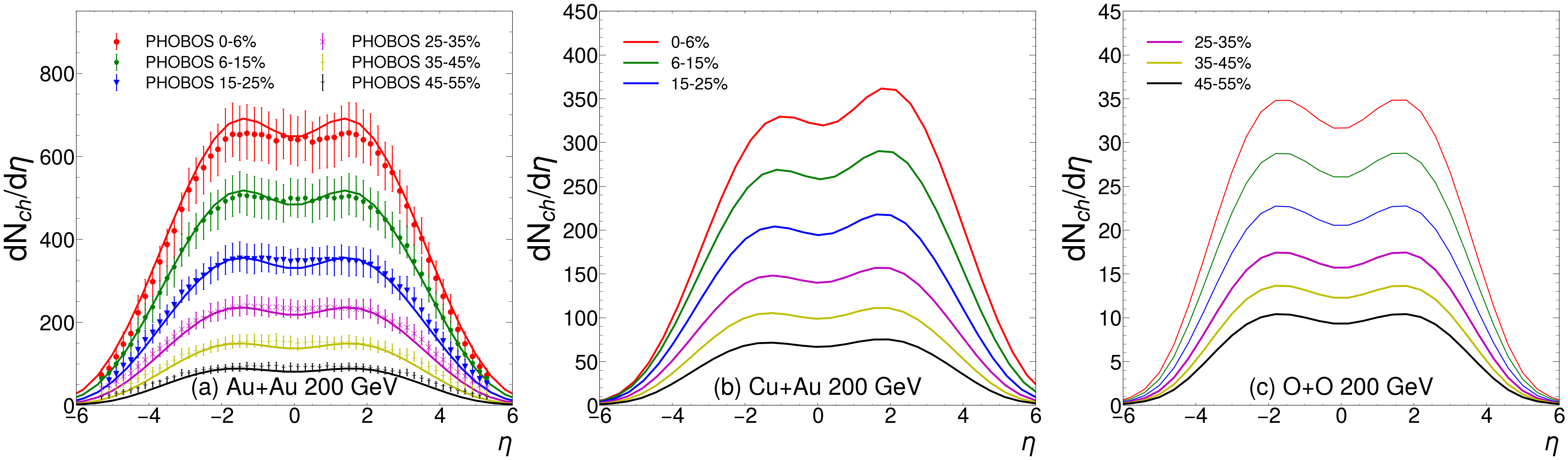}
  \caption{(Color online) The multiplicities of charged hadrons in different centrality bins as functions of pseudorapidity in (a) Au+Au, (b) Cu+Au, and (c) O+O collisions at $\sqrt{s_\mathrm{NN}}=200$~GeV. Results for Au+Au collisions are compared to the PHOBOS data~\cite{PHOBOS:2005zhy}.}
  \label{fig:fig1}
\end{figure*}

\subsection{Hydrodynamic evolution}
\label{section2-2}

With the initial condition constructed in the previous subsection, we use the (3+1)-D viscous hydrodynamics model CLVisc~\cite{Pang:2018zzo,Wu:2021fjf} to simulate the evolution of the QGP event-by-event. The hydrodynamic equations are based on the energy-momentum conservation and the net baryon number conservation as 
\begin{align}
    \partial_\mu T^{\mu\nu} &= 0,\\
    \partial_\mu J^{\mu} &= 0,
\label{eq:15-16}
\end{align}
where $T^{\mu\nu}$ and $J^{\mu}$ are the energy-momentum tensor and the net baryon current respectively. They can be further decomposed as:
\begin{align}
    T^{\mu\nu} &= e U^{\mu}U^{\nu} - P \Delta^{\mu\nu} + \pi^{\mu \nu},\\
    J^\mu &= nU^\mu + V^\mu, 
\label{eq:17-18}
\end{align}
where $e$ is the energy density, $U^{\mu}$ is the 4-velocity of the fluid cell, $P$ is the pressure, $\Delta^{\mu\nu} = g^{\mu\nu} - U^{\mu}U^{\nu}$ is the projector operator, $\pi^{\mu\nu}$ is the shear-stress tensor, $n$ is the net baryon number density, and $V^{\mu}$ is the baryon diffusion current. 
We neglect the effects of bulk viscosity in this study for the purpose of accelerating computing speed. Earlier studies indicate the bulk viscosity may reduce the radial flow of the QGP~\cite{Ryu:2015vwa,Ryu:2017qzn}. This is compensated in our work by adjusting the initial proper time $\tau_0$ to reproduce the mean transverse momenta of the final state charged particles. 

Two model parameters embedded in the $\pi^{\mu\nu}$ and $V^\mu$ terms above are the specific shear viscosity $C_{\eta_\mathrm{v}}$ and the baryon diffusion coefficient $\kappa_\mathrm{B}$. They are related to the shear viscosity $\eta_\mathrm{v}$, the baryon chemical potential $\mu_\mathrm{B}$ and the medium temperature $T$ via
\begin{align}
C_{\eta_\mathrm{v}} &= \frac{\eta_v T}{e+P},\\
\kappa_\mathrm{B} &= \frac{C_\mathrm{B}}{T}n\left[\frac{1}{3}\cot\left(\frac{\mu_\mathrm{B}}{T}\right)-\frac{nT}{e+P}\right].
\end{align}
We set $C_{\eta_\mathrm{v}}=0.08$ and $C_\mathrm{B}=0.4$ through our calculations. The relaxation times are given by $\tau_\pi = 5C_{\eta_\mathrm{v}}/T$ and $\tau_V = C_\mathrm{B}/T$.

We employ the NEOS-BQS equation of state~\cite{Monnai:2019hkn,Monnai:2021kgu} for hydrodynamic evolution, which is based on the lattice QCD calculation at high temperature and vanishing net baryon density, and extended to finite net baryon density using the Taylor expansion. At lower energy density, it transits into the equation of state of hadron gas via a smooth crossover. Detailed discussions on the CLVisc model can be found in Ref.~\cite{Wu:2021fjf}.

\subsection{Hadronization and afterburner}
\label{section2-3}

When the QCD matter becomes sufficiently dilute, quark and gluon degrees of freedom would become confined into hadrons again, and the hydrodynamic description of the bulk evolution should be switched to a transport description of hadronic rescatterings.

On the hadronization hypersurface (chosen as energy density $e_\mathrm{frz}=0.26$~GeV/fm$^3$ in this work), we use the Cooper-Frye formula to obtain the distributions of different hadrons with respect to transverse momentum ($p_\mathrm{T}$), azimuthal angle ($\phi$) and rapidity ($Y$) as:
\begin{equation}
    \frac{dN_h}{p_\mathrm{T}dp_\mathrm{T}d\phi dY}= \frac{g_h}{(2\pi)^3} \int_{\Sigma} p^\mu d \Sigma_\mu f_\mathrm{eq}(1 + \delta f_\pi + \delta f_V),
\label{eq:22}
\end{equation}
where $h$ denotes the hadron species, $g_h$ is its spin degeneracy factor, and $d\Sigma_\mu$ is the 3-D hypersurface element inside the 4-D spacetime determined by the Cornelius routine~\cite{Nabi:2012edo}. The thermal equilibrium distribution $f_\mathrm{eq}(x,p)$ and its out-of-equilibrium corrections $\delta f_\pi(x,p)$ and $\delta f_V(x,p)$ can be calculated using thermal quantities from the hydrodynamic model as~\cite{McNelis:2021acu}:
\begin{align}
    f_\mathrm{eq} &= \left[\exp \left( \frac{p_\mu U^{\mu} - B\mu_\mathrm{B}}{T_\mathrm{f}}\right) \pm 1 \right]^{-1},\\
    \delta f_\pi &= [1 \pm f_\mathrm{eq}(x,p)] \frac{p_\mu p_\nu \pi^{\mu\nu}}{2T_\mathrm{f}^2(e+P)},\\
    \delta f_V &= [1 \pm f_\mathrm{eq}(x,p)] \left(\frac{n}{e+P} - \frac{B}{U^\mu p_\mu}\right)\frac{p^\mu V_\mu}{\kappa_\mathrm{B}/\tau_V}.
\label{eq:19-21}
\end{align}
Here, $T_\mathrm{f}$ represents the chemical freeze-out temperature corresponding to the $e_\mathrm{frz}$ we use, and $B$ is the baryon number of identified particle. 

Based on the Cooper-Frye formalism above, we use the Monte-Carlo method to sample hadrons out of the QGP medium and then feed them into the SMASH~\cite{SMASH:2016zqf,Schafer:2019edr,Mohs:2019iee,Hammelmann:2019vwd,Mohs:2020awg} model for simulating their subsequent scatterings in the hadronic phase. 
SMASH solves the relativistic Boltzmann equation that includes processes of elastic collisions, resonance excitations, string excitations, and decays for hadrons with masses up to about 2~GeV. To improve statistical accuracy, for each hydrodynamic event, we repeat the particle sampling and SMASH simulation 2000 times for Au+Au and Cu+Au collisions. For O+O collisions with smaller sizes and therefore stronger statistical fluctuations, this number is increased to 20000 to ensure statistically stable results.

%%%%% Sec 3

\section{Numerical results}
\label{section3}

\subsection{Particle multiplicity}
\label{section3-1}

\begin{figure*}[tbp!]
  \centering
  \includegraphics[width=0.96\textwidth]{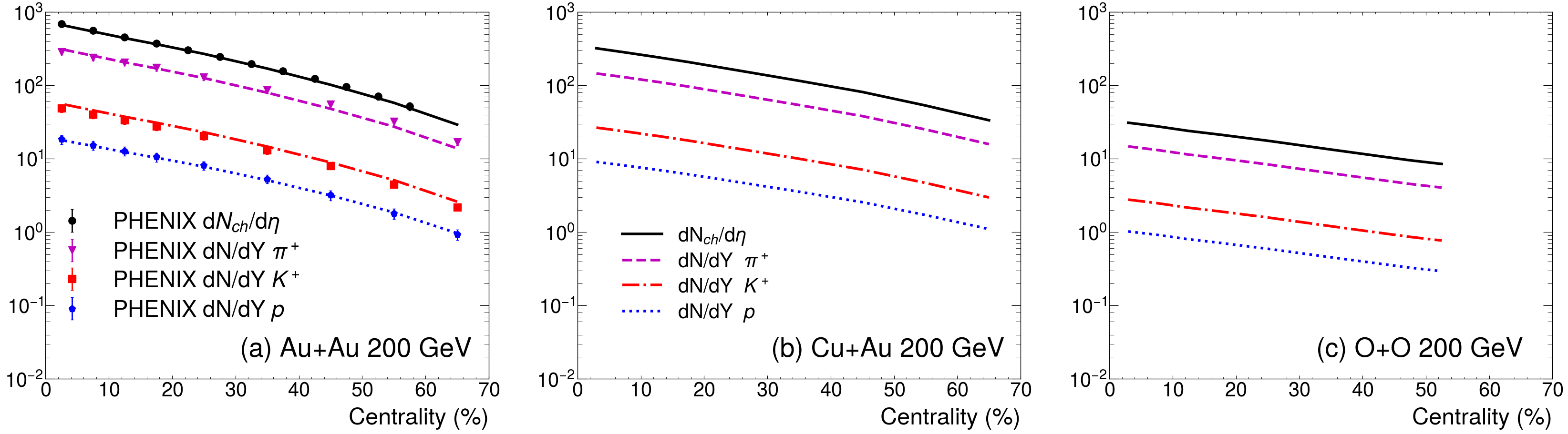}
  \caption{(Color online) The multiplicities of charged hadrons and identified particles as functions of centrality in (a) Au+Au, (b) Cu+Au, and (c) O+O collisions at $\sqrt{s_\mathrm{NN}}=200$~GeV. Results for Au+Au collisions are compared to the PHENIX data~\cite{PHENIX:2003iij,PHENIX:2015tbb}.}
  \label{fig:fig2}
\end{figure*}

We start with validating our hydrodynamic calculation with the charged hadron yields per unit pseudorapidity ($dN_\mathrm{ch}/d\eta$) as functions of pseudorapidity ($\eta$) in Fig.~\ref{fig:fig1}. As shown in panel~(a), with the model setup and parameter tuning presented in the previous section, the hydrodynamic calculation can quantitatively describe both the centrality dependence and the pseudorapidity dependence of charged particle production measured by the PHOBOS Collaboration~\cite{PHOBOS:2005zhy} in Au+Au collisions at $\sqrt{s_\mathrm{NN}}=200$~GeV. 
 
In panels (b) and (c), we present the same calculation for Cu+Au and O+O collisions respectively. The centrality classes are still set according to the previous PHOBOS data. It is important to note that for the asymmetric Cu+Au collisions, the center-of-mass of collisions in our computational frame (located at $Y=0$) needs to be shifted towards the Au-moving ($+\hat{z}$) direction by~\cite{STAR:2017ykf}
\begin{equation}
    y_\mathrm{CM} \approx \frac{1}{2} \ln ( N_\mathrm{part}^\mathrm{Au} / N_\mathrm{part}^\mathrm{Cu} ),
\label{eq:23}
\end{equation}
in order to compare to experimental data. Here, $N_\mathrm{part}^\mathrm{Au}$ and $N_\mathrm{part}^\mathrm{Cu}$ are the average numbers of participant nucleons from Au and Cu nuclei respectively, as evaluated from the MC-Glauber model. This is why we observe a slight shift of each curve's saddle point toward the Au-moving direction in panel (b). This shift becomes more prominent when the imbalance between the energy deposition from Au and Cu nuclei becomes stronger, i.e., in more central collisions. Comparing among the three panels, one can clearly see a decrease in the particle yield as the system size becomes smaller. 

We further present in Fig.~\ref{fig:fig2} the multiplicities of both charged hadrons and identified particles as functions of centrality in the three collision systems. The values of $dN_\mathrm{ch}/d\eta$ (for charged hadrons) and $dN/dY$ (for identified particles) here are taken from their mid-(pseudo)rapidity regions. As shown in panel (a), our model calculation provides a good description of the currently available data from the PHENIX Collaboration on charged hadrons, pions, kaons, and protons in Au+Au collisions~\cite{PHENIX:2003iij,PHENIX:2015tbb}. Predictions for Cu+Au and O+O collisions are provided in panels (b) and (c) respectively. Same as the observation from Fig.~\ref{fig:fig1}, the particle yield decreases from central to peripheral collisions, and also from Au+Au to Cu+Au and then to O+O collisions. Our prediction on the particle yields in O+O collisions based on the Glauber initial condition is comparable to that based on the more advanced IP-Glasma model~\cite{Schenke:2020mbo}.

\subsection{Mean transverse momentum}
\label{section3-2}

\begin{figure*}[tbp!]
  \centering
  \includegraphics[width=0.96\textwidth]{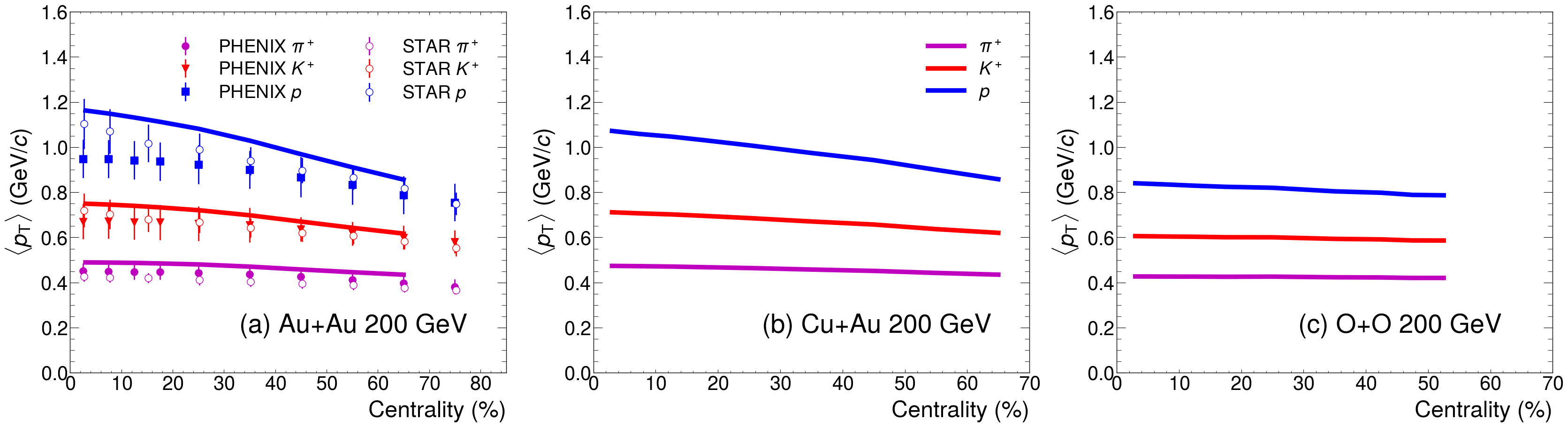}
  \caption{(Color online) The mean transverse momenta of identified particles as functions of centrality in (a) Au+Au, (b) Cu+Au, and (c) O+O collisions at $\sqrt{s_\mathrm{NN}}=200$~GeV. Results for Au+Au collisions are compared to the PHENIX data~\cite{PHENIX:2003iij} and the STAR data~\cite{STAR:2008med}.}
  \label{fig:fig3}
\end{figure*}

The mean transverse momenta $\langle p_\mathrm{T} \rangle$ of final state hadrons help quantify the thermal properties of the QGP and its radial flow developed from hydrodynamic expansion. They are strongly correlated with the $p_\mathrm{T}$ spectra of hadrons~\cite{Pratt:2015zsa,Sangaline:2015isa}, and meanwhile, own the advantage of a better visualization on a linear scale than the $p_\mathrm{T}$ spectra that are conventionally plotted on a logarithmic scale.

Shown in Fig.~\ref{fig:fig3} are the $\langle p_\mathrm{T} \rangle$'s of identified particles as functions of centrality in Au+Au, Cu+Au, and O+O collisions at $\sqrt{s_\mathrm{NN}}=200$~GeV. As seen in panel (a), our calculation reasonably describes the PHENIX~\cite{PHENIX:2003iij} and STAR~\cite{STAR:2008med} data in Au+Au collisions. Nevertheless, there is a hint of overestimation of the proton $\langle p_\mathrm{T} \rangle$ here, possibly due to the neglect of bulk viscosity in hydrodynamic evolution in our current work. This may lead to an underestimate of the proton $v_2$ later. Calculations using the IP-Glasma or Trento initial condition models and taking into account the bulk viscosity can be found in Refs.~\cite{Schenke:2019ruo,Schenke:2020mbo,JETSCAPE:2020mzn}.
The mass ordering in $\langle p_\mathrm{T} \rangle$ can be clearly observed in the figure: being produced from the same QGP medium, heavier hadrons acquire larger $p_\mathrm{T}$ from the thermal background than lighter hadrons do. Meanwhile, compared to lighter hadrons, it is easier for heavier hadrons to gain additional $p_\mathrm{T}$ from the radial flow of the medium, which decreases as centrality increases. Therefore, the $\langle p_\mathrm{T} \rangle$ of heavier hadrons has a stronger dependence on the centrality of heavy-ion collisions. Comparing among the three panels, we also observe a weaker centrality dependence of the hadron $\langle p_\mathrm{T} \rangle$ in O+O collisions than in Au+Au and Cu+Au collisions. This results from the weak radial flow developed in the small-size O+O system, even in its central collisions. In peripheral collisions, the same species of hadrons exhibit similar magnitudes of $\langle p_\mathrm{T} \rangle$ across different collision systems, since the effect of radial flow is negligible in peripheral collisions and the $\langle p_\mathrm{T} \rangle$ is mainly determined by the hadronization temperature of the QCD medium at $\sqrt{s_\mathrm{NN}}=200$~GeV. Note that this $\langle p_\mathrm{T} \rangle$ may depend on beam energy of heavy-ion collisions~\cite{Wu:2021fjf,Shen:2020jwv,Du:2023efk} considering the varying boundary between QGP and hadron gas in the QCD phase diagram as $\sqrt{s_\mathrm{NN}}$ changes.

\subsection{Fluctuations of collective flow}
\label{section3-3}

As discussed in the introduction section, to avoid the difficulty in determining the event plane in realistic experimental measurements, the multi-particle correlation method is usually preferred in evaluating the collective flow coefficients in heavy-ion collisions. 

Consider $m$ is an positive integer, the $n$-th order $2m$-particle azimuthal correlator is defined as~\cite{Borghini:2001vi}
\begin{equation}
\label{eq:def2mCorr}
\langle\langle 2m \rangle\rangle = \langle\langle e^{in\sum_{j=1}^m(\phi_{2j-1}-\phi_{2j})}\rangle\rangle,
\end{equation}
where the inner layer of angle bracket denotes an average over all possible combinations of particles within an event, and the outer layer denotes an average across different events. The corresponding 2-, 4-, and 6-particle cumulants are then given by
\begin{align}
    c_n\{2\} &= \langle\langle 2 \rangle\rangle, \label{eq:cn2}\\
    c_n\{4\} &= \langle\langle 4 \rangle\rangle - 2  \langle\langle 2 \rangle\rangle^2, \label{eq:cn4}\\
    c_n\{6\} &= \langle\langle 6 \rangle\rangle - 9  \langle\langle 4 \rangle\rangle \langle\langle 2 \rangle\rangle +12 \langle\langle 2 \rangle\rangle^3; \label{eq:cn6}
\end{align}
with the 2-, 4-, and 6-particle harmonic coefficients given by
\begin{align}
    v_n\{2\} &= \sqrt{c_n\{2\}}, \\
    v_n\{4\} &= \sqrt[\leftroot{-2}\uproot{2}4]{-c_n\{4\}}, \\
    v_n\{6\} &= \sqrt[\leftroot{-2}\uproot{2}6]{c_n\{6\}/4}. \label{eq:vn6}
\end{align}

By assuming Gaussian distributions of the collective flow fluctuations, one would obtain~\cite{Voloshin:2008dg,Bhalerao:2011yg}
\begin{align}
    v_n\{2\} &\approx \langle v_n\rangle + \sigma_n^2/(2 \langle v_n\rangle), \label{eq:24}\\
    v_n\{4\} &\approx \langle v_n\rangle - \sigma_n^2/(2 \langle v_n\rangle), \label{eq:25}\\
    v_n\{6\} &\approx \langle v_n\rangle - \sigma_n^2/(2 \langle v_n\rangle),
\label{eq:26}
\end{align}
where $\langle v_n\rangle$ is the magnitude of the average of the $\vec{v}_n$ vector (with its direction denoting the event plane angle) in the transverse plane, and $\sigma_n$ is the Gaussian width of the flow fluctuations. Therefore, one can use the ratio $v_n\{4\}/v_n\{2\}$ or $v_n\{6\}/v_n\{2\}$ to quantify the strength of fluctuation in the collective flow coefficient: larger deviation from one implies stronger fluctuation. Equations~(\ref{eq:24}) to~(\ref{eq:26}) are also valid for non-Gaussian distributions of fluctuations as long as their variances satisfy $\sigma_n \ll \langle v_n\rangle$.

Since a direct evaluation of Eq.~(\ref{eq:def2mCorr}) requires 2$m$ loops over the particle list in each event, which is computationally inefficient when $m$ is large, we adopt the $Q_n$-vector method developed in Ref.~\cite{Bilandzic:2010jr} to compute these correlators. For an event consisting of $M$ particles within a desired kinematic region, $Q_n$ is defined as:
\begin{equation}
    Q_n = \sum_{i=1}^M e^{in\phi_i}.
    \label{eq:27}
\end{equation} 
The single-event-averaged 2-, 4-, and 6-particle azimuthal correlators can then be written as:
\begin{align}
    \langle 2 \rangle &= \frac{\lvert Q_n \rvert^2 -M}{M(M-1)},\\
    \langle 4 \rangle &= \frac{\lvert Q_n \rvert^4 + \lvert Q_{2n} \rvert^2 -2\mathfrak{Re}[Q_{2n}Q_{n}^*Q_{n}^*]}{M(M-1)(M-2)(M-3)} \nonumber\\
                     &- 2\frac{2(M-2)\lvert Q_n \rvert^2 - M(M-3)}{M(M-1)(M-2)(M-3)}, \\
    \langle 6 \rangle & = \frac{|Q_{n}|^{6} + 9 |Q_{2n}|^{2} |Q_{n}|^{2} - 6 \mathfrak{Re}[Q_{2n} |Q_{n}|^{2} (Q_{n}^{*})^2]}{M(M - 1)(M - 2)(M - 3)(M - 4)(M - 5)} \nonumber\\
                     &  + 4\frac{  \mathfrak{Re}[Q_{3n} (Q_{n}^{*})^3] - 3  \mathfrak{Re}[Q_{3n} Q_{2n}^{*} Q_{n}^{*}]}{M(M - 1)(M - 2)(M - 3)(M - 4)(M - 5)} \nonumber \\
                     &  + 2\frac{  9(M - 4)  \mathfrak{Re}[Q_{2n} (Q_{n}^{*})^2] + 2  |Q_{3n}|^{2}}{M(M - 1)(M - 2)(M - 3)(M - 4)(M - 5)} \nonumber\\
                     &  - 9\frac{  |Q_{n}|^{4} + |Q_{2n}|^{2}}{M(M - 1)(M - 2)(M - 3)(M - 5)} \nonumber\\
                     &  + 18\frac{  |Q_{n}|^{2} }{M(M - 1)(M - 3)(M - 4)} \nonumber\\
                     &  - \frac{ 6 }{(M - 1)(M - 2)(M - 3)}. 
\label{eq:28-30}
\end{align}
We then average these results over different events and substitute them into Eqs.~(\ref{eq:cn2}) to~(\ref{eq:vn6}) to obtain the collective flow coefficients from the cumulant method.

\begin{figure*}[tbp!]
  \centering
  \includegraphics[width=0.92\textwidth]{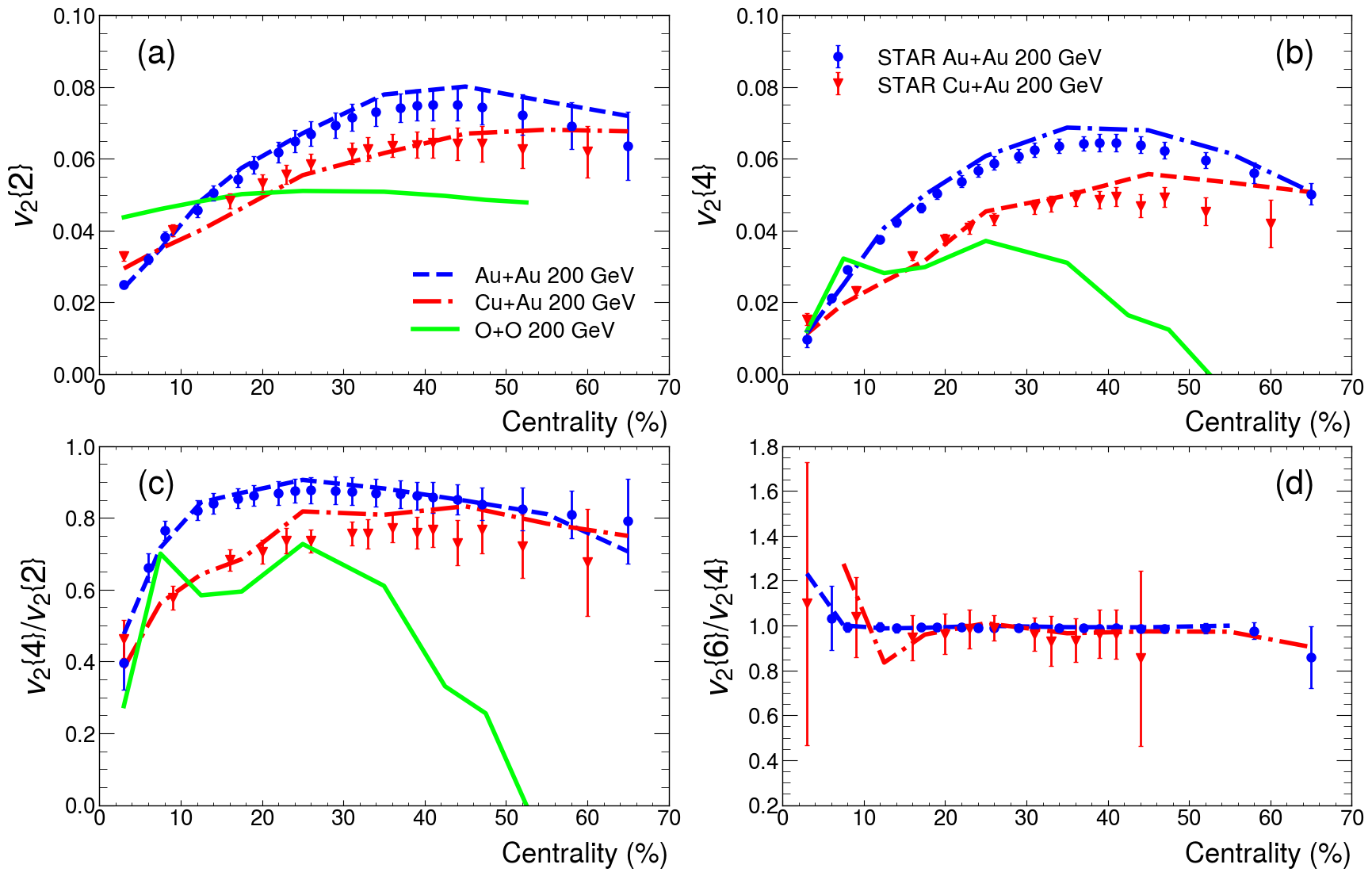}
  \caption{(Color online) The centrality dependence of (a) $v_2\{2\}$, (b) $v_2\{4\}$, (c) $v_2\{4\}$/$v_2\{2\}$, and (d) $v_2\{6\}$/$v_2\{4\}$ of charged particles in Au+Au, Cu+Au, and O+O collisions at $\sqrt{s_\mathrm{NN}}=200$~GeV. Results for Au+Au and Cu+Au collisions are compared to the STAR data~\cite{STAR:2022gki}.}
  \label{fig:fig4}
\end{figure*}

To study the collective flow fluctuations across systems with different shapes and sizes, we implement event-by-event hydrodynamic simulations of Au+Au, Cu+Au, and O+O collisions in this work. In each centrality bin, 500 events are simulated for Au+Au collisions. To enhance statistical accuracy in smaller systems, this number is increased to 1000 for Cu+Au collisions and 1500 for O+O collisions per centrality bin. Additionally, following the STAR Collaboration work~\cite{STAR:2022gki}, we adjust the smallest centrality bin from 0-5\% to 1-5\% for each collision system to avoid possible positive values of $c_n\{4\}$ in very central (0-1\%) collisions, considering that geometric fluctuations in ultra-central collisions can be very strong\cite{Zhou:2018fxx}.

In the upper panels of Fig.~\ref{fig:fig4}, we first present the collective flow coefficients evaluated from the 2-particle [panel (a)] and 4-particle [panel (b)] cumulant methods. The kinematic cuts are chosen as $|\eta|<1$ and $p_\mathrm{T}\in(0.2,4.0)$~GeV/$c$ according to experiment. In general, our calculation provides a reasonable description of the STAR data~\cite{STAR:2022gki} on $v_2\{2\}$ and $v_2\{4\}$ in both Au+Au and Cu+Au collisions at $\sqrt{s_\mathrm{NN}}=200$~GeV, except for some deviation in $v_2\{4\}$ in peripheral Cu+Au collisions. Comparing between these two systems, we observe a slightly larger $v_2$ in Cu+Au than in Au+Au in the most central collisions. This results from stronger initial state fluctuations in a smaller system. On the other hand, at larger centrality where $v_2$ is dominated by the average medium geometry, it is larger in Au+Au than in Cu+Au collisions. We have verified that at centrality greater than 10\%, the eccentricity of Au+Au collisions is larger than that of Cu+Au collisions. Predictions for O+O collisions are also presented here for comparison. Due to the small system size, $v_2$ from O+O collisions is mainly driven by fluctuations, and therefore its value is larger in central collisions while smaller in peripheral collisions compared to those seen in Au+Au and Cu+Au collisions. Since the number of particles produced in peripheral O+O collisions is limited, the statistics of its $v_2\{4\}$ constructed from 4-particle correlation becomes poor at large centrality.

\begin{figure*}[tbp!]
  \centering
  \includegraphics[width=0.97\textwidth]{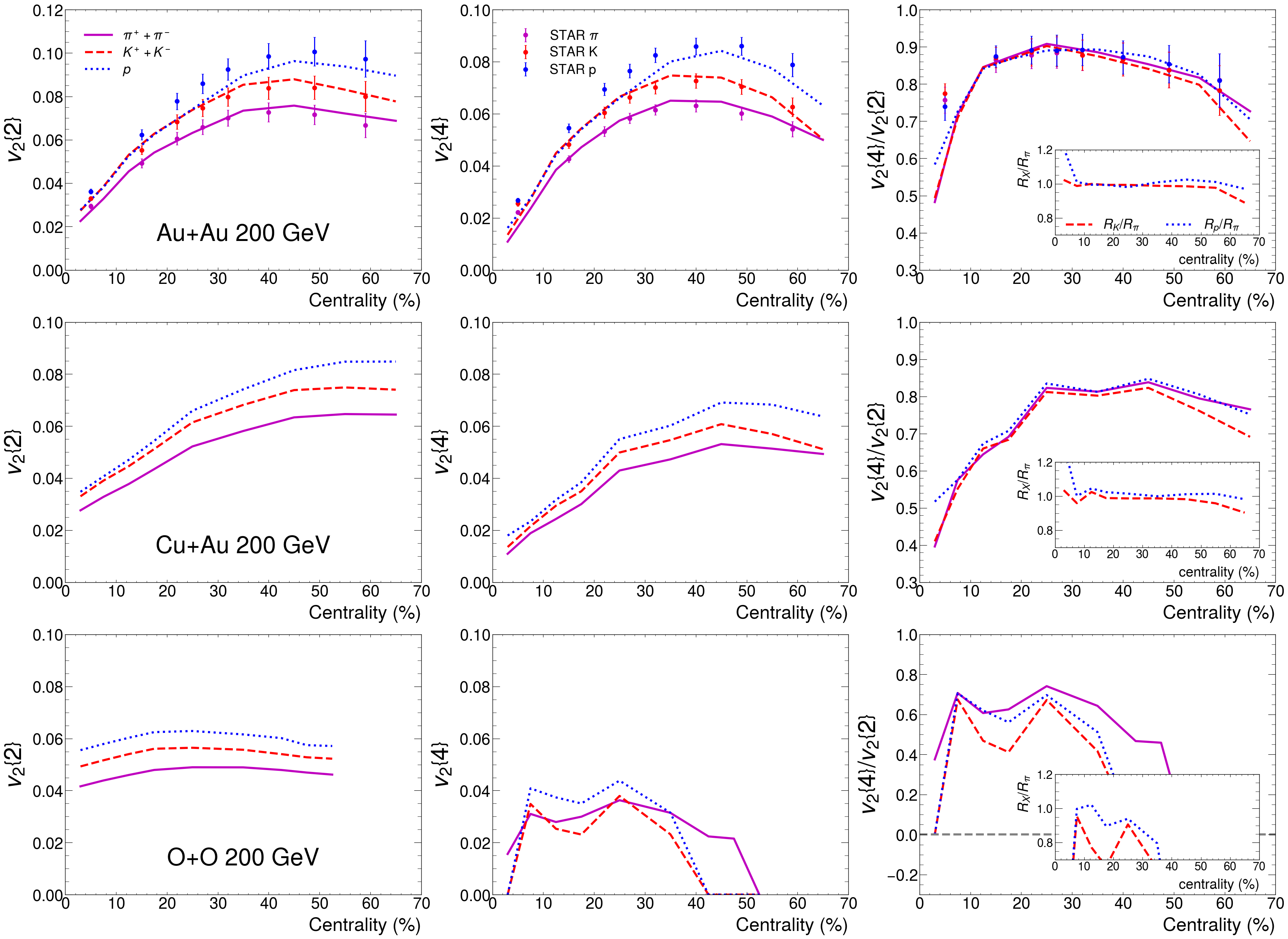}
  \caption{(Color online) The centrality dependence of $v_2\{2\}$ (left column), $v_2\{4\}$ (middle column), and their ratio (right column) of identified particles in Au+Au (upper row), Cu+Au (middle row), and O+O (lower row) collisions at $\sqrt{s_\mathrm{NN}}=200$~GeV. Results for Au+Au collisions are compared to the STAR data~\cite{STAR:2022gki}.}
  \label{fig:fig5}
\end{figure*}

Shown in the lower panels of Fig.~\ref{fig:fig4} are $v_2\{4\}/v_2\{2\}$ [panel (c)] and $v_2\{6\}/v_2\{4\}$ [panel (d)] of charged particles in different collision systems. For Au+Au and Cu+Au collisions, we observe the values of $v_2\{4\}/v_2\{2\}$ are significantly smaller than one in central collisions, indicating the dominant effect of fluctuations on forming $v_2$ in the corresponding region. As centrality increases, this ratio first increases towards one as the average geometry of the medium starts to dominate, and then slightly decreases again as the system becomes small and effect of fluctuations grows again at peripheral collisions. In the mid-central to semi-peripheral region, the eccentricity of the initial state geometry is smaller in Cu+Au than in Au+Au collisions at the same centrality. Meanwhile, fluctuations in the former is stronger due to its smaller system size. As a result, $v_2\{4\}/v_2\{2\}$ is smaller in Cu+Au than in Au+Au collisions within this centrality region. Fluctuations in the small O+O system is strong across the entire centrality region and therefore the corresponding $v_2\{4\}/v_2\{2\}$ keeps significantly below one. Here, we can clearly observe the strong dependence of $v_2\{4\}/v_2\{2\}$ on the size and geometry of the collision system. This is different from the insensitivity of this ratio to the beam energy within the same collision system, as seen in earlier theoretical calculation~\cite{Wu:2021fjf} and experimental data~\cite{STAR:2022gki}. The $v_2\{6\}/v_2\{4\}$ ratios agree with one for both Au+Au and Cu+Au collisions in panel (d). Possible slight deviation from one in very central and very peripheral regions may result from non-Gaussian form of strong fluctuations there. Because of very limited statistics in O+O collisions for constructing $v_2\{6\}$, result for this system is not presented in panel (d).

\begin{figure*}[tbp!]
  \centering
  \includegraphics[width=0.96\textwidth]{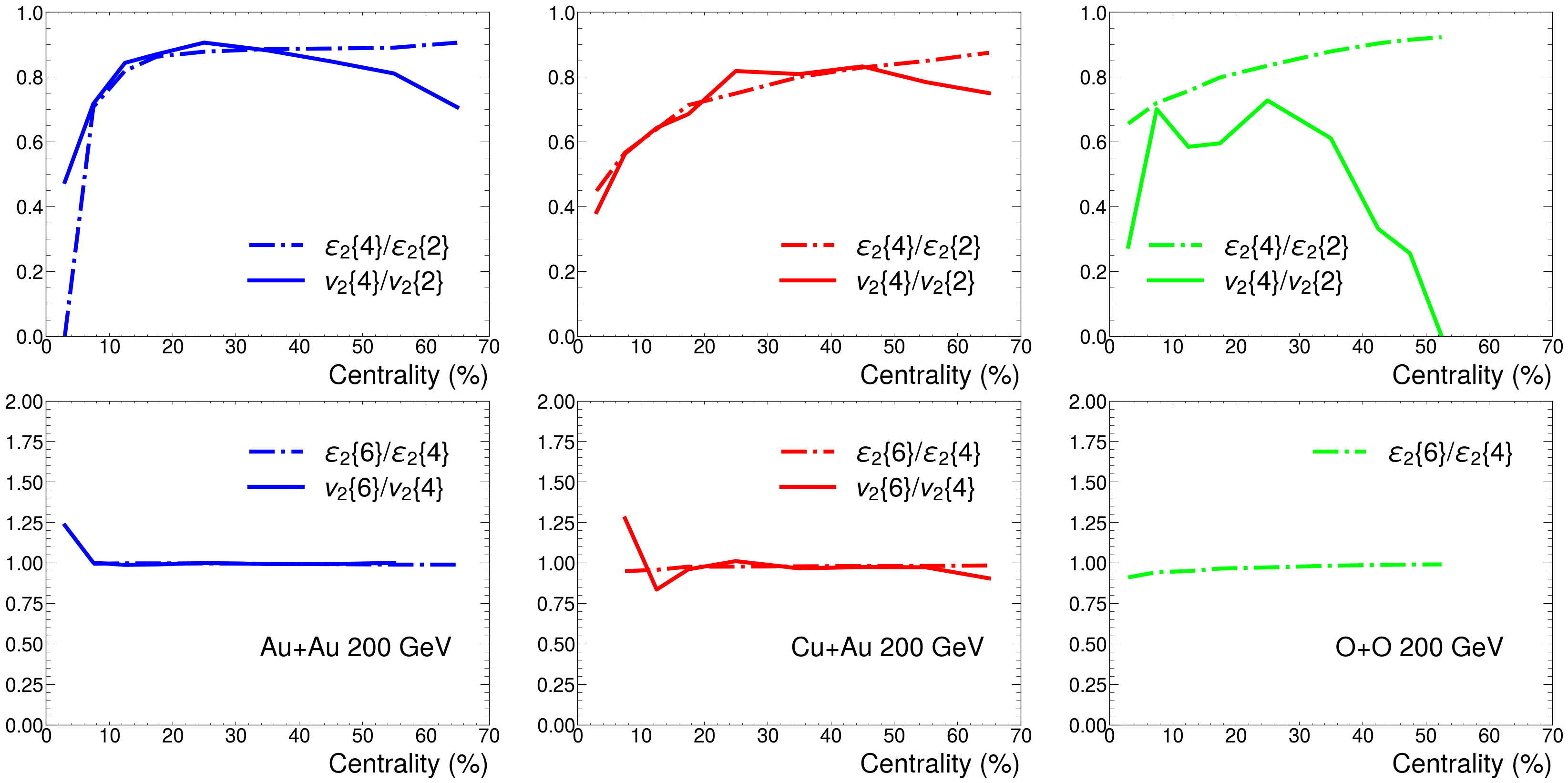}
  \caption{(Color online) Comparison between the centrality dependence of cumulant eccentricity ratios of initial states and cumulant elliptic flow ratios of charged particles in Au+Au (left column), Cu+Au (middle column), and O+O (right column) collisions at $\sqrt{s_\mathrm{NN}} = 200$~GeV; upper row for $\varepsilon_2\{4\}/\varepsilon_2\{2\}$ {\it vs.} $v_2\{4\}/v_2\{2\}$, lower row for $\varepsilon_2\{6\}/\varepsilon_2\{4\}$ {\it vs.} $v_2\{6\}/v_2\{4\}$.}
  \label{fig:fig6}
\end{figure*}

In Fig.~\ref{fig:fig5}, we present $v_2\{2\}$, $v_2\{4\}$, and $v_2\{4\}/v_2\{2\}$ for identified particles in the three collision systems at $\sqrt{s_\mathrm{NN}} = 200$~GeV. Compared to the available data from the STAR Collaboration~\cite{STAR:2022gki}, we provide a good description of $v_2\{2\}$ and $v_2\{4\}$ of pions and kaons, while slightly underestimating them for protons in Au+Au collisions. Since our previous result on the proton $\langle p_\mathrm{T} \rangle$ is close to the upper edges of the data error bars in Fig.~\ref{fig:fig3}~(a), selecting the $p_\mathrm{T}\in(0.2,2.0)$~GeV/$c$ range here, as used in measurements, excludes protons with $p_\mathrm{T}\gtrsim2.0$~GeV/$c$ from our model which can contribute a larger $v_2$ to its average value. It is interesting to note that the mass hierarchy of the $p_\mathrm{T}$-averaged $v_2$ in Au+Au and Cu+Au collisions here -- heavier hadrons show stronger $v_2$ -- is opposite to what one usually sees in the $p_\mathrm{T}$-dependent $v_2$ of identified hadrons below 2~GeV/$c$~\cite{PHENIX:2006dpn,ALICE:2014wao,ALICE:2022zks,STAR:2022ncy}. This is because of the harder $p_\mathrm{T}$ spectra of heavier hadrons which add more weights from the higher $p_\mathrm{T}$ region to their average $v_2$. In other words, one actually compares the $v_2$ of heavier hadrons with higher average $p_\mathrm{T}$ to the $v_2$ of lighter hadrons with lower average $p_\mathrm{T}$ in the $p_\mathrm{T}$-averaged $v_2$ here. Although different species of hadrons show different magnitudes of $v_2\{2\}$ and $v_2\{4\}$, they have similar $v_2\{4\}/v_2\{2\}$ ratios in Au+Au and Cu+Au systems, except in very central and very peripheral collision. This indicates these hadrons keep the memory of the momentum space fluctuations of the QGP, and are not strongly affected by additional fluctuations during the hadronization and hadronic afterburner processes. Contrarily, $v_2\{4\}/v_2\{2\}$ is no longer independent of hadron species in O+O collisions. Although the statistical errors in our results for O+O collisions are large, the dependence of $v_2\{4\}/v_2\{2\}$ on hadron species is visible across a wide centrality range. This dependence can result from the fluctuations in hadronization and hadronic afterburner, which become more prominent for a smaller system. These non-initial-state fluctuations may also be important in very central Au+Au and Cu+Au collisions, where the collective flows are dominated by fluctuations. To better compare $v_2\{4\}/v_2\{2\}$ between different hadron species, we show its ratio between kaons and pions, and between protons and pions in the sub-panels in the last column.

\subsection{Connection to initial state fluctuations}
\label{section3-4}

Earlier studies~\cite{Magdy:2018itt,STAR:2022gki,Alba:2017hhe,Schenke:2019ruo,Magdy:2020gxf,Magdy:2021sba,Rao:2019vgy} suggest that the collective flow fluctuations, quantified by $v_2\{4\}/v_2\{2\}$, predominantly originates from the fluctuations of the initial state geometry in large nuclear collision systems, such as Au+Au and Pb+Pb collisions. To characterize the anisotropy of the initial state geometry, one may evaluate its $n$-th order eccentricity as 
\begin{equation}
    \varepsilon_n = \frac{\sqrt{\langle r^n \cos(n\phi)\rangle^2 + \langle r^n \sin(n\phi)\rangle^2}}{\langle r^n\rangle},
\label{eq:34}
\end{equation}
where $r$ and $\phi$ are the position and azimuthal angle of an element of nuclear matter with respect to its center of mass, and the angle brackets denote average within an event. In this work, both participant nucleons and binary collision points sampled from the MC-Glauber model contribute to the elements in this average, with $\alpha$ as their relative weight as shown in Eq.~(\ref{eq:4}).

Within the hypothesis of linear hydrodynamic response, $v_n \propto \varepsilon_n$, one may construct the cumulants of eccentricities according to Eqs.~(\ref{eq:cn2}) to~(\ref{eq:cn6})~ as~\cite{Qiu:2011iv,Ma:2016hkg}:
\begin{align}
    c_{\varepsilon_n}\{2\} &= \langle\varepsilon_n^2 \rangle,\\
    c_{\varepsilon_n}\{4\} &= \langle\varepsilon_n^4 \rangle - 2\langle\varepsilon_n^2 \rangle^2,\\
    c_{\varepsilon_n}\{6\} &= \langle\varepsilon_n^6 \rangle - 9\langle\varepsilon_n^4 \rangle\langle\varepsilon_n^2 \rangle + 12\langle\varepsilon_n^2 \rangle^3,
\label{eq:36-38}
\end{align}
where the angle brackets denote average over different events. The $n$-th order eccentricities defined by the cumulant method then read:
\begin{align}
    \varepsilon_n\{2\} &= \sqrt{c_{\varepsilon_n}\{2\}},\\
    \varepsilon_n\{4\} &= \sqrt[\leftroot{-2}\uproot{2}4]{-c_{\varepsilon_n}\{4\}},\\
    \varepsilon_n\{6\} &= \sqrt[\leftroot{-2}\uproot{2}6]{c_{\varepsilon_n}\{6\}/4}.
\label{eq:39-41}
\end{align}
Analogous to the collective flow coefficients determined through the cumulant method, the cumulant eccentricities here encode information of fluctuations in the initial state geometry of the QGP. Therefore, similar to using $v_n\{4\}/v_n\{2\}$ to measure the fluctuations in the final state collective flow, one can use $\varepsilon_n\{4\}/\varepsilon_n\{2\}$ to quantify the strength of the initial state fluctuations. 

To study the correlation between the initial state geometric fluctuation and the final state flow fluctuation, we compare $\varepsilon_2\{4\}/\varepsilon_2\{2\}$ and $v_2\{4\}/v_2\{2\}$ in the upper panel of Fig.~\ref{fig:fig6}, and compare $\varepsilon_2\{6\}/\varepsilon_2\{4\}$ and $v_2\{6\}/v_2\{4\}$ in the lower panel, for Au+Au, Cu+Au, and O+O collisions at $\sqrt{s_\mathrm{NN}}=200$~GeV. The cumulant eccentricities $\varepsilon_2\{k\}$ here are obtained from averaging over 50000 MC-Glauber profiles in each centrality bin of each collision system. In the upper three panels, we observe a monotonic increase of $\varepsilon_2\{4\}/\varepsilon_2\{2\}$ for the three systems from central to peripheral collisions. This indicates, in central collisions, the geometric anisotropy in the initial state is mainly contributed by fluctuations, while in peripheral collisions, it is mainly determined by the average shape of the overlapping region between the two nuclei. In Au+Au and Cu+Au collisions, $\varepsilon_2\{4\}/\varepsilon_2\{2\}$ agrees well with $v_2\{4\}/v_2\{2\}$ except in very central and very peripheral regions. This suggests the initial state fluctuations are the main source of the final state collective flow fluctuations in these large enough collision systems. When the centrality is large, the systems become so small that effects of additional fluctuations, e.g. nonlinear hydrodynamic response, hadronization, and hadronic afterburner, become important, leading to smaller values of $v_2\{4\}/v_2\{2\}$ than $\varepsilon_2\{4\}/\varepsilon_2\{2\}$. As revealed in Refs.~\cite{Noronha-Hostler:2015dbi,Giacalone:2017uqx}, nonlinear hydrodynamic response has a significant contribution to the elliptic flow fluctuations in peripheral collisions. For the small O+O system, sources other than the initial state always have strong contributions to the flow fluctuations, and thus $v_2\{4\}/v_2\{2\} < \varepsilon_2\{4\}/\varepsilon_2\{2\}$ for the whole centrality range.

In the lower three panels of Fig.~\ref{fig:fig6}, we observe the values of $\varepsilon_2\{6\}/\varepsilon_2\{4\}$ are close to one in all the three systems. No obvious deviation is seen between $\varepsilon_2\{6\}/\varepsilon_2\{4\}$ and $v_2\{6\}/v_2\{4\}$ in Au+Au and Cu+Au collisions, except for some hints of slight deviation in very central and very peripheral regions. The result of $v_2\{6\}/v_2\{4\}$ in O+O collisions is not available due to the limited statistics of hadrons produced in these small systems.

\section{Summary and Outlook}
\label{section4}

Using the (3+1)-D viscous hydrodynamic model CLVisc coupled to a MC-Glauber initial condition, Cooper-Frye formalism for particlization, and SMASH for hadronic rescatterings, we investigate the yields, mean transverse momenta, elliptic flow fluctuations for both charged hadrons and identified particles in Au+Au, Cu+Au, and O+O collisions at $\sqrt{s_\mathrm{NN}}=200$~GeV. By incorporating contributions from both participant nucleons and binary collisions to the initial entropy density and net baryon density distributions, and assuming the hydrodynamic parameters solely depend on the collision energy, our model calculation provides a satisfactory description of the $dN/d\eta$, $\langle p_\mathrm{T}\rangle$, $v_2\{2\}$, $v_2\{4\}$, and $v_2\{4\}/v_2\{2\}$ data currently available at RHIC, and also provides predictions for Cu+Au and O+O systems where full measurements are yet to be done.

Comparing across the three systems, we find the particle yields significantly increase with the system size of nuclear collisions. While a clear mass hierarchy of $\langle p_\mathrm{T} \rangle$ exists in all systems, the centrality dependence of $\langle p_\mathrm{T} \rangle$ becomes weaker in a smaller system due to a weaker radial flow developed from hydrodynamic expansion. Comparing between Au+Au and Cu+Au collisions, we see the elliptic flow in very central collisions is larger in the latter system due to stronger initial state fluctuations in a smaller system. On the other hand, it is larger in the former system at large centrality due to the larger eccentricity of the overlapping region in Au+Au than in Cu+Au collisions. Elliptic flow from the small O+O system is mainly driven by fluctuations, and therefore appears larger than that in the other two systems at small centrality, while smaller at large centrality. We quantify the fluctuation strength of $v_2$ using the ratio of its values estimated from the 4-particle and 2-particle cumulant methods, and find this $v_2\{4\}/v_2\{2\}$ ratio is far below one in the most central Au+Au and Cu+Au collisions, and then increases towards one as centrality increases, but decreases again at very large centrality. This indicates strong $v_2$ fluctuations in the most central and peripheral collisions, while less fluctuations in mid-central to semi-peripheral collisions. The value of $v_2\{4\}/v_2\{2\}$ is always much less than one for O+O collisions, signifying strong fluctuations in small systems across all centralities. The value of $v_2\{6\}/v_2\{4\}$ is generally consistent with one in Au+Au and Cu+Au collisions, though slight deviation may happen in the most central and very peripheral regions due to possible non-Gaussian forms of strong fluctuations there.

By comparing $v_2\{4\}/v_2\{2\}$ between different species of hadrons, and comparing this final state $v_2\{4\}/v_2\{2\}$ with the initial state $\varepsilon_2\{4\}/\varepsilon_2\{2\}$, we can gain insights on different sources of collective flow fluctuations in different collision systems. We find that in Au+Au and Cu+Au collisions, $v_2\{4\}/v_2\{2\}$ is almost species independent, except in very central and very peripheral collisions. On the contrary, $v_2\{4\}/v_2\{2\}$ clearly depends on the hadron species in O+O collisions. These findings suggest that in large collision systems, hadrons can well preserve the fluctuations inherited from the QGP, while in small collision systems, additional fluctuations from hadronization and hadronic afterburner have a non-negligible impact on the collective flows of hadrons. These additional fluctuations may also affect the most central collisions of large systems, where collective flow originates from fluctuations. Similar conclusions can also be drawn from the comparison between $v_2\{4\}/v_2\{2\}$ and $\varepsilon_2\{4\}/\varepsilon_2\{2\}$ in different systems, where we find the strength of the collective flow fluctuations $v_2\{4\}/v_2\{2\}$ follows that of the initial geometric fluctuations $\varepsilon_2\{4\}/\varepsilon_2\{2\}$ in mid-central to semi-peripheral Au+Au and Cu+Au collisions, while $v_2\{4\}/v_2\{2\}<\varepsilon_2\{4\}/\varepsilon_2\{2\}$ in peripheral Au+Au and Cu+Au collisions, and in O+O collisions over the entire centrality region. The deviation between $v_2\{4\}/v_2\{2\}$ and $\varepsilon_2\{4\}/\varepsilon_2\{2\}$ can also be attributed to the nonlinear hydrodynamic response during the QGP expansion. Therefore, we anticipate future measurements on $v_2\{4\}/v_2\{2\}$ of identified particles in different collision systems can not only help quantify the system size and shape dependences of the strength of collective flow fluctuations, but also shed light on the origins of these fluctuations in different systems.

While this work helps improve our quantitative understanding on the dependences of collective flow fluctuations on the size and shape of nuclear systems in relativistic heavy-ion collisions, it can be further improved in several directions.  
Instead of a manual adjustment of our model parameters, the Bayesian statistical analysis method~\cite{Bernhard:2016tnd,JETSCAPE:2020avt} needs to be introduced for calibrating our model, which is necessary for drawing a more precise conclusion on the strength of collective flow fluctuations and their possible origins. Along our current investigation on effects of the average initial shape of nuclear matter on the flow fluctuations, another promising topic is using these observables to image deformed nuclei in the initial state~\cite{Zhang:2021kxj,Jia:2021qyu,Bally:2022vgo}, and connecting our current study to the isobar experiments at RHIC~\cite{Sinha:2022cvj,Bairathi:2023chw}. Furthermore, hydrodynamic fluctuations~\cite{Young:2014pka,Sakai:2020pjw,Sakai:2021rug,Kuroki:2023ebq} and non-Gaussian form of fluctuations~\cite{CMS:2017glf,An:2022jgc} are also exciting topics for a more delicate understanding of the fluctuation phenomena in heavy-ion collisions. These aspects will be explored in our future efforts.

\begin{acknowledgements}
We are grateful for valuable discussions with Jian Deng and Li Yan. This work was supported by the National Natural Science Foundation of China (NSFC) under Grant Nos. 12175122, 2021-867, 12225503, 11890710, 11890711, and 11935007. X.-Y.W. was also supported in part by the Natural Sciences and Engineering Research Council of Canada, and in part by US National Science Foundation (NSF) under Grant No. OAC-2004571.
\end{acknowledgements}

\bibliographystyle{h-physrev5}
\bibliography{reference}

\end{document}